\documentclass[usenatbib]{mn2e}
\usepackage{epsfig}
\def\nei{\hbox{Ne\,{\sc i}}}
\newcommand{\vsini}{\mbox{$v\sin\!i$}}
\newcommand{\logg}{\mbox{$\log g$}}
\newcommand{\teff}{\mbox{$T_{\mathrm{eff}}$}}

\begin{document}

\title[HgMn stars]{Neon abundances in normal late-B and mercury-manganese 
stars}

\author[M. M. Dworetsky and J. Budaj]
{M. M. Dworetsky$^1$\thanks{Guest Observer, Lick
Observatory} and J. Budaj$^{1,2}$\thanks{E-mail: (MMD)
mmd@star.ucl.ac.uk (JB) budaj@ta3.sk}\\
$^1$Department of Physics and Astronomy, University College London, Gower
Street, London, WC1E 6BT \\
$^2$Astronomical Institute of the Slovak
Academy of Sciences, 05960 Tatransk\'{a} Lomnica, Slovak Republic}

\maketitle

\begin{abstract} 

We make new Non-LTE calculations to deduce abundances of neon from
visible-region \'{e}chelle spectra of selected \nei\ lines in 7 normal
stars and 20 HgMn stars.  We find that the best strong blend-free Ne line
which can be used at the lower end of the \teff\ range is $\lambda$6402,
although several other potentially useful \nei\ lines are found in the red
region of the spectra of these stars.  The mean neon abundance in the
normal stars ($\log A =8.10$) is in excellent agreement with the standard
abundance of neon (8.08).  However, in HgMn stars, neon is almost
universally underabundant, ranging from marginal deficits of 0.1-0.3 dex
to underabundances of an order of magnitude or more.  In many cases, the
lines are so weak that only upper limits can be established.  The most
extreme example found is $\upsilon$ Her with an underabundance of at least
1.5 dex. These underabundances are qualitatively expected from radiative
acceleration calculations, which show that Ne has a very small radiative
acceleration in the photosphere and is expected to undergo gravitational
settling if mixing processes are sufficiently weak, and there is no strong
stellar wind.  According to the theoretical predictions of
\citet{landolvau}, the low Ne abundances place an important constraint on
the intensity of such stellar winds, which must be less than
$10^{-14}M_{\sun}$\,yr$^{-1}$ if they are non-turbulent.

\end{abstract}


\section{Introduction} 
\label{section:intro}
HgMn stars are a subclass of chemically peculiar star occupying the
spectral region corresponding approximately to MK types B9--B6
(10\,500--16\,000\,K).  Due to low helium abundances, their spectral
classes are generally placed by observers in the A0--B8 range.
Observationally, they are characterized by extremely low rotational
velocities, weak or non-detectable magnetic fields and photometric
variability, and atmospheric deficiencies of light elements (e.g., He, Al,
N) coupled with enhancements of the heavy elements (e.g., Hg, Mn, Pt, Sr,
Ga).  In addition, the heavy elements also have non-terrestrial isotopic
abundances (\citealt{kcs97}; \citealt*{bohl}). The currently favoured
mechanism for explaining these anomalies is the radiative diffusion
hypothesis \citep{m70}.  This work has been advanced in the form of a
parameter-free model \citep{m86}.

The quiescent atmospheres of these stars makes them one of the best
natural laboratories for studying the competing processes of gravitational
settling and radiative levitation \citep{vau82}. In the absence of
disrupting mechanisms such as convection, rotationally-induced meridional
currents, high microturbulence and magnetic fields, certain rare elements
can reach a factor of 10$^5$ enhancement over their standard abundance.
Because of the strength and sharpness of normally exotic spectroscopic
lines, HgMn stars are also useful for constraining fundamental atomic data
\citep{l95}.

Although there have been many studies of individual HgMn stars and of the
abundances of many elements across a sample of HgMn stars, we have been
unable to find any papers mentioning the abundance of Ne in HgMn stars,
with the recent exception of a paper by \citet{adel00}, in which LTE
calculations established that \nei\ line strengths in $\kappa$ Cnc implied
overabundance of 0.64 dex, while an underabundance was found in HR\,7245.  
The He-wk star 3~Cen~A also seemed to be overabundant in Ne relative to
the Sun (we will look again at these results in Section~\ref{disc}).
However, Ne is known to exhibit strong non-LTE effects in B stars,
according to the original investigations by \citet{auer73}, who showed
that Non-LTE (NLTE) methods yield \nei\ lines that are nearly double the
strength expected from LTE calculations at \teff\ = 15\,000\,K.  Thus the
overabundances reported may well be due to neglect of NLTE considerations.

The lack of Ne observations in late B stars is slightly surprising,
because Ne is an important and interesting element. Its standard abundance
(\citealt{ag89}; \citealt*{grev96}), which was deduced from the solar
wind, nebular spectroscopy, and stellar observations and analyses such as
those of \citeauthor{auer73}, is comparable with that of C, N and O. It is
interesting also because its atomic structure resembles that of He with
very high first ionization potential (about 22\,eV). Consequently, all its
resonance lines and the ground state photoionization continuum are in the
Lyman continuum where the stellar energy flux in the photosphere of late B
stars is low.  One would then expect that radiative acceleration on Ne may
not be enough to balance gravity and Ne should sink. Indeed, theoretical
calculations by \citet*{landolvau}, who considered a nonturbulent mass
loss or stellar wind in the stellar envelopes, predict that there will be:
(1) neon underabundances if the mass loss rate is less than
$10^{-14}M_{\sun}$\,yr$^{-1}$; (2) neon overabundances for mass loss in
the range $10^{-14}-10^{-12}M_{\sun}$\,yr$^{-1}$; or (3) normal neon
abundances for mass loss over $10^{-12}\,M_{\sun}$\,yr$^{-1}$. Also, our
own calculations of radiative accelerations in the atmospheres
\citep{budw00} predict a pattern of general photospheric underabundances
for Ne.

In this paper, we present an abundance analysis of visible \nei\ lines
based on a full NLTE treatment of the strength of $\lambda$6402. It is the
strongest unblended line of \nei\ in the visible spectrum.  We also
demonstrate that, to a close order of magnitude, several other \nei\
lines tend to have similar NLTE enhancements and
can be used if spectra showing $\lambda$6402 are not available.  We find
that for most HgMn stars, neon turns out to be underabundant, suggesting
that the first scenario of \citet{landolvau} is the most likely one.

\section{Observations}
\label{section:obs}

Our stellar sample is based upon that of \citet{sd93}, who analysed IUE
data on the UV resonance lines of iron-peak elements in 26 HgMn, 4
superficially normal and 10 normal stars. We observed definite detections
or determined upper limits for \nei\ in 21 of the HgMn stars in the
\citet{sd93} sample and 11 of the normal and superficially normal group.  
Some of the other stars in the two samples were lacking data in the red
region or were cooler than 10\,000\,K and we did not expect to observe any
\nei\ lines.  Physical parameters of the stars in this study are given in
Table~\ref{table:stars}.

\begin{table*}
\caption{Programme stars: basic data and adopted atmospheric parameters.}
\begin{tabular}{llllrllllr}
\hline
Star &HD &Spectral type &V$_r$ &\teff &\logg &$\xi$ &Ref.
&\vsini &Ref. \\
 & & &(km\,s$^-$$^1$) &(K) &(dex\,cm\,s$^-$$^2$) &(km\,s$^-$$^1$) &
&(km\,s$^-$$^1$) \\ \hline \multicolumn{10}{c}{\it {Normal and
superficially normal stars}} \\ $\pi$\,Cet &17081 &B7\,V &+15\,SB &13\,250
&3.80 &0.0 &[1] &25 &[3] \\ 134\,Tau &38899 &B9\,IV &+18\,V &10\,850 &4.10
&1.6 &[1] &30 &[3] \\ $\tau$\,Her &147394 &B5\,IV&$-$14\,V? &15\,000 &3.95
&0.0 &[3] &32 &[3] \\ $\zeta$\,Dra &155763 &B6\,III&$-$17\,V? &12\,900
&3.90&2.5: & &34 &[3] \\ $\alpha$\,Lyr &172167 &A0\,Va &$-$14\,V &9\,450
&4.00 &2.0 &[4] &24 &[3] \\ HR\,7098 &174567 &A0\,Vs &$-$\ 3 &10\,200
&3.55 &1.0 &[3] &11 &[3] \\ 21\,Aql &179761 &B8\,II-III &$-$\ 5\,V
&13\,000 &3.50 &0.2 &[1] &17 &[3] \\ HR\,7338 &181470 &A0\,III &$-$14\,SBO
&10\,250 &3.75 &0.5 &[3] &\ 3 &[3] \\ $\nu$\,Cap &193432 &B9.5\,V &$-$\
2\,V? &10\,300 &3.90 &1.6 &[1] &27 &[3] \\ HR\,7878 &196426 &B8\,IIIp
&$-$23 &13\,050 &3.85 &1.0: & &\ 6 &[9] \\ 21\,Peg &209459 &B9.5\,V &$-$\
0 &10\,450 &3.50 &0.5 &[3] &\ 4 &[3] \\ \multicolumn{10}{c}{\it {HgMn
stars}} \\ 87\,Psc &7374 &B8\,III &$-$16\,V &13\,150 &4.00 &1.5 &[3] &21.0
&[5] \\ 53\,Tau &27295 &B9\,IV &+12\,SBO &12\,000 &4.25 &0.0 &[2] &\ 6.5
&[5] \\ $\mu$\,Lep &33904 &B9\,IIIpHgMn &+28 &12\,800 &3.85 &0.0 &[2]
&15.5 &[5] \\ HR\,1800 &35548 &B9\,pHgSi &$-$\ 9\,V? &11\,050 &3.80 &0.5
&[3] &\ 3.0 &[5] \\ 33\,Gem &49606 &B7\,III &+13 &14\,400 &3.85 &0.5: &
&22.0 &[5] \\ HR\,2676 &53929 &B9.5\,III &+\ 6\,V? &14\,050 &3.60 &1.0: &
&25.0 &[5] \\ HR\,2844 &58661 &B9\,pHgMn &+21\,V &13\,460 &3.80 &0.5: &
&27.0 &[5] \\ $\nu$\,Cnc &77350 &A0\,pSi &$-$15\,SBO &10\,400 &3.60 &0.1
&[6] &13 &[6] \\ $\kappa$\,Cnc &78316 &B8\,IIIpMn &+24\,SB1O &13\,500
&3.80 &0.0 &[2] &\ 7 &[5] \\ HR\,4072 &89822 &A0\,pSiSr:Hg: &$-$\ 0\,SB2O
&10\,500 &3.95 &1.0 &[7] &\ 3.2 &[7] \\ $\chi$\,Lup &141556 &B9\,IV &+\
5\,SB2O &10\,750 &4.00 &0.0 &[11] &\ 2.0 &[7] \\ $\iota$\,CrB &143807
&A0\,p:Hg: &$-$19\,SB &11\,000 &4.00 &0.2 &[6] &\ 1.0 &[7] \\
$\upsilon$\,Her & 144206 &B9\,III &+\ 3 &12\,000 &3.80 &0.6 &[1] &\ 9.0
&[5] \\ $\phi$\,Her &145389 &B9\,p:Mn: &$-$16\,SB1O &11\,650 &4.00 &0.4
&[8] &10.1 &[5] \\ HR\,6997 &172044 &B8\,II-IIIpHg &$-26$\,SBO &14\,500
&3.90 &1.5 &[3] & 36.0 &[5] \\ 112\,Her &174933 &B9\,II-IIIpHg
&$-$20\,SB2O &13\,100 &4.10 &0.0: & &\ 5.5 &[12] \\ HR\,7143 &175640
&B9\,III &$-$26\,V? & 12\,100 &4.00 &1.0 &[3] &\ 2.0 &[5] \\ HR\,7361
&182308 &B9\,IVpHgMn &$-$20 V? &13\,650 &3.55 &0.0 &[3] &\ 8.2 &[5] \\
46\,Aql &186122 &B9\,IIIpHgMn &$-$32 &13\,000 &3.65 &0.0 &[3] &\ 3.0 &[5]
\\ HR\,7664 &190229 &B9\,pHgMn &$-$22\,SB1 &13\,200 &3.60 &0.8 &[8] &\ 8.0
&[5] \\ HR\,7775 &193452 &A0\,III &$-$18$^a$ &10\,800 &3.95 &0.0 &[3] &\
0.8 &[10] \\ \hline \end{tabular} \begin{minipage}{15.1cm} $^a$Hoffleit \&
Warren (1991) cite HR\,7775 as SB1O although this is a confusion with
$\beta$\,Cap (HR\,7776). \\ \ \\ Notes: Spectral types and radial velocity
data are from \citet{bsc91}. Values of \teff\ and \logg\ are from
\citet{sd93}, or [12] in the case of 112\,Her. Values of `V' and `V?',
respectively, indicate known or suspected radial velocity variables; `SB':
spectroscopic binary (`SB1' and `SB2', respectively, denote single- and
double-lined systems); `O': published orbit (see \citealt*{batten89}).
Microturbulence parameters $\xi$ appended by a colon (`:') are approximate
and were derived solely from UV Fe\,II lines by \citet{sd93}. \\
\\
References: [1] \citet{adfuhr85}, [2] \citet{adel88a}; [3]
\citet{kcs92}; [4] \citet{gigas}; [5] \citet*{djs98};
[6] \citet{ad89}; [7] \citet{harman}; [8] \citet{adel88b}; [9]
\citet{cow80}; [10] \citet*{bohl}; [11] \citet*{wahl94};
[12] \citet*{r96}.

\end{minipage}

\label{table:stars}
\end{table*}

All observations were obtained with the Hamilton \'{E}chelle Spectrograph
\citep[HES;][]{vogt} at Lick Observatory, fed by the 0.6-m Coud\'{e}
Auxilliary Telescope (CAT), during four runs in 1994--1997. Further
details of the instrument can be found in \citet{mischb}.  Shortly before
our observations in 1994, some of the HES optical components were
replaced, improving the resolution and instrumental profile, and making it
possible to use the full field of the 2048 $\times$ 2048 CCDs to maximum
advantage.  We used both the unthinned phosphor-coated Orbit CCD (Dewar
13) and from July 1995 the thinned Ford CCD (Dewar 6), depending on
availability as the latter was shared with the multi-object spectrograph
on the 3-m telescope.  The spectral range for the observations was
3800--9000\,\AA. Typical signal-to-noise (S/N) per pixel in the centres of
orders ranged from 75 to 250.  The Orbit CCD is cosmetically very clean,
with very few bad pixels or columns, while the thinned Ford CCD contains
several column defects but offers a much higher detector quantum
efficiency in the blue.  We used the Ford CCD whenever it was available.
With the slit settings used, the combination of spectrographs and CCDs
gave resolutions $R\!\approx\!46\,500$. Flat fields were made using the
polar axis quartz lamp and wavelength calibrations were obtained with a
Th--Ar comparison.

The \'{e}chelle spectra were extracted and calibrated using standard {\sc
iraf} extraction packages \citep{churchill,valdes}, running on UCL's
Starlink node.  Previous measurements \citep{csa98} showed that there were
no measurable effects of parasitic light (residual scattered light)  in
the line profiles provided that general scattered light in the adjacent
interorder spaces was taken as the subtracted background.  In practice the
residual scattered light was less than approximately 1 percent; we have
therefore made no corrections for it. Allen's method is based on a direct
comparison of the solar spectrum (as reflected from the roof of the CAT
coelostat) observed with the HES, with the Kitt Peak Solar Flux Atlas
\citep{kpno}.  As the latter was obtained using a Fourier Transform
Spectrometer, it has no measurable parasitic light.  The Kitt Peak
spectrum is convolved with a suitable instrumental profile to match the
HES data; both spectra must be normalized at the same points for a valid
comparison.  The ratio of summed equivalent widths of various features
with good adjacent continuum points, in many different spectral orders,
provides the measure of the amount of parasitic light.

\section{Abundance determination}
\label{section:abunds}

\subsection{Stellar parameters and stellar atmospheres}
\label{sub:params}

Effective temperatures and surface gravities of programme stars are
summarized in Table~\ref{table:stars}.  In general, the parameters adopted
follow our previous work \citep*{sd93,djs98,jda99}.  Seven stars are noted
as double-lined spectroscopic binaries in Table~\ref{table:ewidths}; one
of these (HR\,1800) is better described as a close visual binary in which
we can see evidence of the secondary spectrum as rotationally-broadened
features. The parameters and light ratios quoted in all seven cases are
those adopted for the primary star.  Suitable light ratios for other
wavelength regions were found by the use of \citet{cd13} model atmosphere
fluxes.  The light ratio estimated this way for $\lambda$6402 is given in
Table~\ref{table:binary}, where it is representative of the values
throughout the range $\lambda\lambda$5800-6700. The adopted light ratio
for the visual binary HR\,1800 ($\rho = 0.243$ arcsec) is 2.45, based on
$\Delta H_p = 0.96$ mag from {\it The Hipparcos Catalogue}
\citep{ESA1997}.  Another star, 33 Gem, is suspected of being double-lined
but there is not yet any information on the orbit or light ratio
\citep{hl93}; we treat it as a single star or `average component.' We note
that \citet*{ad96} also treated 33 Gem as a single star, noting that the
question of binarity could not be conclusively resolved with their data.

\begin{table*}
\caption{Binary stars: adopted stellar data and light ratios}
\begin{tabular}{lcccccr}
\hline
Star &$\lambda$ &L$_A$/L$_B$ &$\teff_A/\logg_A$ &$\teff_B/\logg_B$
&L$_A$/L$_B$ &Ref.\\
 &(\AA) & &(K)/(cgs)  &(K)/(cgs)  &6402\AA & \\
\hline
HR\,7338 &4481 &3.16 &10\,250/3.8 &8\,500/4.0 &2.72 & [1] \\
HR\,1800 &$H_p$ &2.45 &11\,050/3.8 &9\,500/4.0 &2.34 & [2] \\
$\kappa$\,Cnc &5480 &11.5 &13\,200/3.7 &8\,500/4.0 &10.70  & [3]
\\
HR\,4072 &4520 &5.45 &10\,650/3.8 &8\,800/4.2 &5.01 & [4] \\
$\chi$\,Lup &4520 &3.65 &10\,650/3.9 &9\,200/4.2 &3.35 & [4] \\
$\iota$\,CrB &4520 &2.70 &11\,000/4.0 &9\,000/4.3 &2.46 & [4] \\
112\,Her &4520 &6.20 &13\,100/4.1 &8\,500/4.2 &5.20 & [5] \\
\hline
\end{tabular}

\begin{minipage}{10.4cm}
Note: The entry for HR\,1800 is the ratio quoted for the broadband $H_p$
filter, which we assume to be the light ratio at H$\beta$.
References: [1] \citet{petrie}; [2] \citet{ESA1997}; [3] \citet{r98}; [4]
\citet{harman} and \citet*{jda99}; [5] \citet{r96}.
\end{minipage}
\label{table:binary}
\end{table*}

\subsection{Atomic data for the LTE approximation}
\label{sub:atomic}

As all the Ne lines in the stars observed are either weak or, except for
the hottest stars such as $\tau$ Her, not strongly saturated, the main
atomic parameter of critical importance for LTE calculations is the
oscillator strength, given as $\log gf$ in Table~\ref{table:eqwid}.  We
take our oscillator strengths from the calculations of \citet{mjs}, who
showed that his calculations were in excellent agreement (within 10 per
cent) with other recent theoretical and laboratory data such as that of
\citet{hartmetz} for the 3s-3p transitions of interest in this work, and
also in excellent agreement with the critically evaluated $gf$-values as
given by \citet{auer73}.

For the radiative damping, we assumed the classical damping constant
$\Gamma_{\rm R} = 2.223\times10^{7}/\lambda^2$ s$^{-1}$ [$\lambda$ in
$\mu$m]. This is a good approximation (within a factor of 2) for these
lines as the typical lifetime of the upper levels is about 20 ns, and the
abundances are not sensitive to the adopted values in any event. Van der
Waals contributions to line broadening are also expected to be very small;
a suitable approximation by \citet{W67} was used.  For Stark broadening,
we adopted the recent experimental results of \citet*{delval}, and used an
estimate of the temperature scaling factor proportional to $T^{0.4}$ to
convert their $w_m$ at 18\,000\,K to values at 12\,000\,K, by multiplying
by an average factor of 0.85 \citep{griem74}.  One line, $\lambda$5852,
was not included in their list and we adopted the simple approximation
given in CD23 data (\citealt{cd23}).  In general, the measured values
which we used are about 3 times the values in CD23 for lines in common. We
carried out worst-case sensitivity tests by varying the
\citeauthor{delval} Stark broadening by a factor of 2 for the strongest
lines in $\tau$ Her; the largest effect on derived abundances was less
than 0.01 dex.

\subsection{Equivalent widths and LTE results}
\label{sub:ew_LTEabunds}

Estimated abundances for several identified \nei\ lines were determined
using the exact curve-of-growth technique in the LTE approximation. We
measured the equivalent widths, $W_{\lambda}$, of Ne absorption lines in
the programme spectra by numerical integration in the {\sc dipso v3.5}
package \citep{dipso} and compared them to the calculated values for each
line, which were generated by our spectrum-synthesis code {\sc uclsyn}
\citep{sd88,kcs92}. The necessary atmospheric parameters given in
Table~\ref{table:stars} -- \teff, \logg, and microturbulence ($\xi$) --
were taken from \citet{sd93}, except for 112\,Her where we used the values
given by \citet*{r96}.  In the cases of the seven binaries with double
spectra, we adopted the light ratios cited in Section~\ref{sub:params} in
order to correct for dilution effects.  The equivalent widths (corrected
for binarity where necessary) and LTE abundances for several lines are
given in Table~\ref{table:ewidths}. We used the 2\,km\,s$^{-1}$ grid of
\citet{cd13} models, interpolating to produce a model at the chosen \teff\
and \logg\ of each star.

We searched a list of \nei\ lines from \citet*{wsg66} in the range
$\lambda\lambda5800-6800$ and narrowed the list to include only those
which were fairly strong, without evident blending problems, and not
situated at the ends of \'{e}chelle orders where the spectra are noisiest.
In a few cases the quality of the spectra justify quoting equivalent
widths to the nearest 0.1\,m\AA.  For each star a mean LTE abundance,
weighted by equivalent width, was calculated on the scale $\log
N(\rm{H})=12.00$. To investigate the consistency of the results from the
selected lines, the deviations of each line from the mean of all the
lines, $\log (A/A_{\star})$, were calculated for each star where a
meaningful average could be computed.  These are summarized in
Table~\ref{table:ewidths} where the values of $\overline{\log
(A/A_{\star})}$ represent the mean deviations from the overall LTE
abundance for each line.  The small mean deviations imply that the results
for each line are broadly consistent with one another and the relative
$gf$-values of \citet{mjs}.  However, one line ($\lambda$6402) is
considerably stronger than all the others, and is well-suited for
abundance determinations in the largest number of stars, especially those
at the low-\teff\ end of the sequence and with abundances apparently below
the standard value. In the remainder of this paper, we shall consider only
this line, although future investigators may wish to consider some of the
other lines further.

\begin{table*}

\caption{Ne I equivalent widths (m\AA) and LTE abundances for normal and
HgMn programme stars on the scale $\log N ({\rm H}) = 12$}

\label{table:eqwid}
\begin{tabular}{lcccccccccccccc}

\hline 

               &\multicolumn{2}{c}{$\lambda$5852.49}&
                \multicolumn{2}{c}{$\lambda$6096.16}&    
                \multicolumn{2}{c}{$\lambda$6266.50}&
                \multicolumn{2}{c}{$\lambda$6382.99}&
                \multicolumn{2}{c}{$\lambda$6402.25}&
                \multicolumn{2}{c}{$\lambda$6598.95}&
                \multicolumn{2}{c}{$\lambda$6717.04}\\

Star &$W_\lambda$ & $\log A$ &$W_\lambda$ & $\log A$ &$W_\lambda$ & $\log A$
& $W_\lambda$ &$\log A$ & $W_\lambda$&$\log A$ &$W_\lambda$ & $\log A$
& $W_\lambda$ &$\log A$ \\
\hline
 &\multicolumn{13}{c}{\it{Normal and superficially normal stars}}&\\
$\pi$\,Cet      & 8&8.13&10&8.06&15&8.43&18&8.44&39&8.54&--&--&11&8.37\\ 
134\,Tau        &--&--  &--&--  &--&--  &--&--  &$\leq15$&$\leq8.78$
                &--&--  &--&--  \\
$\tau$\,Her     &22&8.52&20&8.25&--&--  &26&8.41&59&8.67&--&--  &30&8.76\\
$\zeta$\,Dra    &13:&8.49: &20:&8.59: &--&--  &24:&8.73:&30&8.32 &--&--
&--&-- \\
$\alpha$\,Lyr   &--&--  &--&--  &--&--  &--&--  &$<10 $&$<9.18$&--&--&--\\
HR\,7098        &--&--  &--&--  &--&--  &--&--  &1.8:&7.55:&--&--&--&-- \\
21\,Aql         &10&8.21&12&8.12& 9&8.07&12&8.12&35&8.38&--&--  &13&8.43\\
HR\,7338$^a$    &--&--&--&--&--&--&--&--&4.0&8.05&--&--&--&--\\
$\nu$\,Cap      &--&--&--&--&--&--&--&--&$\leq10$&$\leq8.69$&--&--&--&--
\\
HR\,7878        & 8&8.18&14&8.33& 8&8.10&16&8.42&30&8.34& 4&7.90&10&8.38\\
21\,Peg&5&8.65&4&8.37&$\leq4$&$\leq8.52$&$\leq2$&$\leq8.03$&9&8.36&
$\leq2$&$\leq8.40$&$\leq2$&$\leq8.35$\\[1mm]
 &\multicolumn{13}{c}{\it{HgMn stars}}&\\
87\,Psc 	&--&--&--&--&--&--&--&--&$\leq8$&$\leq7.41$&--&--&--&--\\
53\,Tau         &--&--&--&--&--&--&--&--&$\leq5$&$\leq7.63$&--&--&--&--\\
$\mu$\,Lep 	&--&--&--&--&--&--&--&--&$\leq10$&$\leq7.60$&--&--&--&--\\
HR\,1800$^a$    &--&--&--&--&--&--&--&--&$\leq7$&$\leq8.02$
&--&--&--&--\\
33\,Gem		&19&8.47&27&8.52&12&8.08&14&8.04&38&8.22&15&8.34&17&8.41\\
HR\,2676        &--&--&--&--&--&--&--&--&25&7.83&--&--&--&--\\
HR\,2844        &--&--&--&--&--&--&--&--&$\leq16$&$\leq7.71$&--&--&--&--\\
$\nu$\,Cnc      &--&--&--&--&--&--&--&--&$\leq5$&$\leq8.05$&--&--&--&--\\
$\kappa$\,Cnc$^a$ 
&21&8.69&22&8.53&22&8.65&21&8.49&38&8.44&12&8.38&20&8.71\\
HR\,4072$^a$    &--&--&--&--&--&--&--&--&4.1&8.04&--&--&--&--\\
$\chi$\,Lup$^a$ &--&--&--&--&--&--&--&--&4.1&7.95&--&--&--&--\\
$\iota$\,CrB$^a$&--&--&--&--&--&--&--&--&3.7&7.76&--&--&--&--\\
$\upsilon$\,Her
&--&--&--&--&--&--&--&--&$\leq1.0$&$\leq6.63$&--&--&--&--\\
$\phi$\,Her 	&--&--&--&--&--&--&--&--&2.0&7.17&--&--&--&--\\
HR\,6997        &15&8.30&19&8.24&20:&8.38:&20&8.25&37&8.15&--&--&--&--\\
112\,Her$^a$    &2.4&7.64&--&--&1.8&7.43&--&--&7.1&7.42&$\leq1.2$
&$\leq7.43$&$\leq3.6$&$\leq7.91$\\
HR\,7143        &--&--&$\leq1:$&$\leq7.3:$&$\leq1:$&$\leq7.4:$
&$\leq2:$&$\leq7.6:$&$\leq3.0$&$\leq7.21$&$\leq1:$&$\leq7.6:$&--&--\\
HR\,7361   	&18&8.47&23&8.44&18&8.38&20&8.33&40&8.35&13&8.30&--&--\\
46\,Aql         &$\leq2$&$\leq7.4$&2:&7.2:&3:&7.4:&3:&7.4:
&9&7.41&2:&7.5:&$\leq2$&$\leq7.4$\\
HR\,7664	&8&8.07&13&8.16&11&8.17&12&8.10&29&8.16&8&8.13&10&8.24\\
HR\,7775        &$\leq3$&$\leq8.42$&$\leq2$&$\leq8.05$&$\leq2$&$\leq8.22$
&6&8.72&--&--&$\leq1$&$\leq8.16$&$\leq4$&$\leq8.79$\\
\\
$\overline{\log (A/A_{\star})}$&& -0.02&&
-0.05&&-0.05&&-0.04&&-0.02&&-0.13&&+0.10\\\\
$\log gf$ &&-0.49&&-0.31&&-0.37&&-0.24&&+0.33&&-0.35&&-0.35\\

\hline
\end{tabular}
\begin{minipage}{18cm}
$^a$Binaries with two spectra.  The $W_\lambda$ values are corrected for
dilution effects as described in the text.  Colons (`:') indicate
uncertain values.
\end{minipage}
\label{table:ewidths}
\end{table*}

\subsection{A weak blending line?}
\label{blend}

Although we have chosen our list of Ne lines to be as blend-free as
possible, there is a predicted weak blending feature in the CD23 list
\citep{cd23} adjacent to the important line \nei\ $\lambda$6402.246:
Fe\,{\sc ii} at $\lambda$6402.397. In programme stars
(Table~\ref{table:stars}) with approximately standard or lower Fe
abundances, this line would have a typical strength of about 0.5~m\AA, too
small to affect our results in any significant way.  However, in the few
stars with enhanced Fe abundances such as 112\,Her \citep[][$\log A({\rm
Fe})=8.40$]{sd93}, the strongest iron-rich case, the possible blending
effect would have raised the apparent abundance of Ne by 0.3\,dex.  It
should be noted that for sharp-lined stars, the displacement of the blend
is enough to make its existence apparent.  The existence of this line with
the $gf$-value listed remains to be confirmed; future work directed at
refining the neon analysis should address the question of its actual
strength with better observations.

\section{NLTE calculations and abundances}
\label{section:NLTE}
\subsection{Calculations and \nei\ model atom}
\label{sub:calcsNLTE}

The first full NLTE calculations of neon line strengths for \nei\ were
made by \citet{auer73} using NLTE model atmospheres.  Unfortunately for
our purposes, their analysis was restricted to the hotter stars with
$\teff>15000$\,K.  Recently, their calculations were revisited and
extended to cooler temperatures by \citet{sigut}. \citeauthor{sigut} used
the $T-\tau$ relations, particle densities, and electron number densities
from \citeauthor{cd13} LTE line-blanketed models and solved the restricted
NLTE problem, i.e. only the equations of radiative tranfer and statistical
equlibrium for Ne. His grid of equivalent widths also has rather large
steps for our purposes: 2000\,K, 0.5\,dex, 0.5\,dex in temperature,
gravity and Ne abundance, respectively.

In this section we examine in detail the temperature, gravity and
Ne-abundance region where all our HgMn stars are found, solve the full
NLTE problem using NLTE model atmospheres, and find a convenient way to
represent the NLTE effects in the \nei\ $\lambda$6402 line so that
straightforward interpolation via LTE models can be done.  For the
calculation of NLTE atmosphere models and level populations we used the
{\sc {tlusty195}} code described in more detail in \citet{hub88}, and in
\citet{hula92,hula95}. Here H\,{\sc{i}} and \nei\ were treated as explicit
ions, which means that their level populations were calculated in NLTE and
their opacity was considered. Other elements like He, C, N, and O were
allowed to contribute to the particle and electron number density in LTE.
Synthetic spectra and equivalent widths were then calculated with the {\sc
{synspec42}} code \citep*{hulajef95}. In the following, if not stated
otherwise, `in LTE' means `in LTE considering the NLTE model of the
atmosphere'.
    
It is not possible to list here all the input parameters entering the NLTE
calculations and above mentioned codes. We will mention only those which
are most crucial for this particular problem or are different from those
which could be generated interactively, e.g. by the very useful interface
tool {\sc modion} \citep{varosi}, part of the {\sc tlusty} package for
creating the model of the atom from the TOPbase data. We provide a copy of
our input model for {\sc tlusty}, which can be downloaded by anyone who
wishes to repeat the calculations (Dworetsky \& Budaj 2000).

\begin{table}
\caption{\nei\ energy levels considered. Column 1: the Paschen
level designation, col. 2: the $nlpqr$ notation of Seaton (1998),
col. 3: ionization energy in cm$^{-1}$, col. 4: statistical weight of the
level.}
\begin{center}
\begin{tabular}{llll}
\hline
Paschen & Seaton  &Energy    & g  \\
\hline
$2p_6$ 1S  &2p   &  174192.4 &  1. \\    
$1s_5$     &3s332&   40148.6 &  5. \\   
$1s_4$     &3s331&   39731.2 &  3. \\   
$1s_3$     &3s110&   39371.8 &  1. \\   
$1s_2$     &3s111&   38301.7 &  3. \\   
$2p_{10}$  &3p311&   25932.7 &  3. \\   
$2p_9$     &3p353&   24533.4 &  7. \\   
$2p_8$     &3p352&   24366.2 &  5. \\   
$2p_7$     &3p331&   24068.8 &  3. \\   
$2p_6$     &3p332&   23874.6 &  5. \\   
$2p_5$     &3p131&   23418.3 &  3. \\   
$2p_4$     &3p132&   23331.9 &  5. \\   
$2p_3$     &3p310&   23273.0 &  1. \\   
$2p_2$     &3p111&   23152.0 &  3. \\   
$2p_1$     &3p110&   21219.7 &  1. \\   
$2s_5$     &4s332&   15589.3 &  5. \\   
$2s_4$     &4s331&   15394.4 &  3. \\   
$2s_3$     &4s110&   14810.5 &  1. \\   
$2s_2$     &4s111&   14655.8 &  3. \\   
$3d_6$     &3d310&   12680.8 &  1. \\   
$3d_5$     &3d311&   12666.3 &  3. \\   
$3d_4'$    &3d374&   12600.1 &  9. \\   
$3d_4$     &3d373&   12598.3 &  7. \\   
$3d_3$     &3d332&   12583.2 &  5. \\   
$3d_2$     &3d331&   12553.8 &  3. \\   
$3d_1''$   &3d352&   12490.8 &  5. \\   
$3d_1'$    &3d353&   12489.0 &  7. \\   
$3s_1''''$ &3d152&   11781.8 &  5. \\   
$3s_1'''$  &3d153&   11780.3 &  7. \\   
$3s_1''$   &3d132&   11770.5 &  5. \\   
$3s_1'$    &3d131&   11754.8 &  3. \\   
\hline
\end{tabular}
\end{center}
\label{table:levels}
\end{table}

The spectrum of \nei\ is that of an inert gas where the LS-coupling breaks
down and terms and multiplets do not provide an appropriate description of
the atom, so that at least the lower terms and multiplets must be split
into individual levels and transitions. We considered explicitly the first
31 levels of \nei\ as \citet{auer73}, plus continuum, with each of the
levels treated separately (Table \ref{table:levels}). Paschen designations
and experimental energies for the levels were taken from \citet{moore}.
Photoionization cross-sections for the terms were taken from the TOPbase
data base \citep{cunto}, as calculated by \citet{hibb}.  We fit individual
photoionization cross-sections to about 10-15 points using the {\sc
modion} code.  It was assumed that the photoionization cross-section is
the same for the term as for the level but it was scaled to the particular
level threshold. We used the calculated oscillator strengths of
\citet{mjs} as before.  For collisional excitation rates we used the van
Regemorter formula as in \citet{auer73}. For collisional ionization we
used equation (5.79) of \citet{mihalas78} with $\bar{g}_{i}=0.1$. With
this input data we calculated the \nei\ $\lambda$6402 equivalent width for
the same abundance ($10^{-4}$), $f$-value (0.431), microturbulence $\xi$
(4\,km\,s$^{-1}$) and similar H-He NLTE models ($\teff=15000, 20000, \log
g=4$) as \citeauthor{auer73} to check and compare our calculations.  (For
the hotter model, two terms of Ne\,{\sc ii} plus continuum were also
considered explicitly.) The results are listed in
Table~\ref{table:auer_compare} and are in very close agreement. To compare
our results with those by \citet{sigut}, and to check the calculations for
lower temperatures, we solved the similar {\em restricted} NLTE problem
using \citet{cd13} CD13 LTE line blanketed models (computed with $\xi=2$
km\,s$^{-1}$) with our \nei\ atom model plus Hubeny's H\,{\sc{i}} atom
model (9 explicit levels plus continuum) and the same Ne abundance ($1.12
\times 10^{-4}$), $f$-value (0.428), and $\xi$ (5\,km\,s$^{-1}$) as
\citeauthor{sigut}. The results are compared in
Table~\ref{table:sigut_compare} and are also in very good agreement.\\

\begin{table} 
\caption{Equivalent widths of \nei\ $\lambda$6402 (m\AA) for two
models, comparing this work (D+B) with \citet{auer73}.} 
\begin{center}
\begin{tabular}{ccccc} 
\hline
 \teff\  & \multicolumn   {2}{c}{15000} & \multicolumn{2}{c}{20000} \\
\hline
   &LTE    & NLTE &  LTE & NLTE \\
\hline
D+B   &    30 & 46  &    39 &  74\\
A+M   &    28 & 45  &    40 &  79\\
\hline
\end{tabular}
\end{center}
\label{table:auer_compare}
\end{table}

\begin{table}
\caption{Equivalent widths of \nei\ $\lambda$6402 (m\AA) for two
\citeauthor{cd13} models with $\teff=12000\, \&\, 17000$\,K, $\log g=4$,
$\xi=5$\,km\,s$^{-1}$, comparing this work (D+B) with \citet{sigut}.}
\begin{center}
\begin{tabular}{ccccc}
\hline
 \teff\  & \multicolumn{2}{c}{12000} & \multicolumn{2}{c}{17000} \\
\hline
         & NLTE    & NLTE/LTE &  NLTE & NLTE/LTE \\
\hline
D+B     &    19 & 1.31  &    89 &  1.81\\
Sigut   &    18 & -     &    86 &  1.80\\
\hline
\end{tabular}
\end{center}
\label{table:sigut_compare}
\end{table}

We found that although microturbulence can affect the
equivalent widths due to desaturation effects, for the observed stars
($0 \leq\xi\leq 2.5$) it has negligible effect on
the atmosphere model, \nei\ level populations, and LTE/NLTE equivalent
width ratio, $R = W_{\lambda({\rm LTE})}/W_{\lambda({\rm NLTE})}$.  For
this reason, our task can be considerably simplified as one can calculate
a grid of models and LTE/NLTE ratios for only one $\xi=0$. Also, the NLTE
effect of varying the Ne abundance is quite small. While decreasing the Ne
abundance from the standard value (8.08) by 1.0 dex reduces the equivalent
width considerably, it raises the corresponding LTE/NLTE ratio $R$ by only
$0.03\pm0.01$. Consequently, the main results can be gathered into
Table~\ref{table:ratios} listing $R(\teff, \logg)$ for standard Ne
abundance and zero microturbulence.  The ratio $R$ ranges from around
$0.6-0.7$ at the high \teff\ end of the HgMn domain to nearly 1.0 at the
cool end of the sequence (where the lines of neon disappear and can no
longer be studied).  \citet{dwbu2000} provide a short Fortran77 code to
interpolate in the grid (including the small effects of abundance) for
anyone who wishes to undertake their own interpolations for $R$.

\begin{table}
\caption{LTE/NLTE equivalent width ratio $R$ (\nei\ $\lambda$6402).}
\begin{center}
\begin{tabular}{crllll}
\hline
\teff &\logg =  & 3.50 & 3.75  &  4.00 & 4.25 \\
\hline
11000        &    & 0.78 & 0.81  & 0.84 & 0.87\\
12000        &    & 0.72 & 0.75  & 0.79 & 0.82\\
13000        &    & 0.67 & 0.70  & 0.74 & 0.77\\
14000        &    & 0.63 & 0.66  & 0.69 & 0.73\\
15000        &    & 0.59 & 0.62  & 0.65 & 0.69\\
\hline
\end{tabular}
\end{center}
\label{table:ratios}
\end{table}

\citet{hub81} pointed out that apart from hydrogen, C\,{\sc{i}} and
Si\,{\sc{i}} are the other most important ions to be included explicitly
in the NLTE calculations in early A stars.  We have checked that for this
particular problem, including these elements along with He may again
slightly affect the equivalent width, but it does not affect $R$
significantly.  The line strength results were somewhat sensitive to the
collisional excitation.  A more detailed analysis of the current
precision of neon NLTE calculations can be found in \citet{sigut}.

\begin{table}

\caption{Non-LTE abundances from \nei\ $\lambda$6402 for normal and HgMn
programme stars on the scale $\log A ({\rm H}) = 12.00$. Programme stars
with high abundance upper limits or highly uncertain measurements have
been omitted.}

\begin{tabular}{lccccc}

\hline 

Star &\teff&$\logg$&$W_\lambda$&$W_{\rm LTE}/W_{\rm NLTE}$&$\log A$ \\
\hline
\multicolumn{6}{c}{\it{Normal and superficially normal stars}}\\
$\pi$\,Cet      & 13250 & 3.80 & 39 & 0.69 & 8.18 \\ 
$\tau$\,Her     & 15000 & 3.95 & 59 & 0.64 & 8.15 \\
$\zeta$\,Dra    & 12900 & 3.90 & 30 & 0.73 & 8.07 \\
21\,Aql         & 13000 & 3.50 & 35 & 0.67 & 8.01 \\
HR\,7338        & 10250 & 3.75 & 4.0& 0.86 & 8.04 \\
HR\,7878        & 13050 & 3.85 & 30 & 0.71 & 8.04 \\
21\,Peg         & 10450 & 3.50 & 9  & 0.82 & 8.22 \\[1mm]
\multicolumn{6}{c}{\it{HgMn stars}}\\
87\,Psc 	& 13150 & 4.00 & $\leq8$ &0.76 & $\leq7.27$ \\
53\,Tau         & 12000 & 4.25 & $\leq5$ &0.83 & $\leq7.53$ \\
$\mu$\,Lep 	& 12800 & 3.85 & $\leq10$&0.75 & $\leq7.43$ \\
HR\,1800        & 11050 & 3.80 & $\leq7$ &0.83 & $\leq7.90$ \\
33\,Gem		& 14400 & 3.85 & 38 & 0.67 & 7.87 \\
HR\,2676        & 14050 & 3.60 & 25 & 0.66 & 7.54 \\
HR\,2844        & 13460 & 3.80 &$\leq16$&0.72&$\leq7.50$ \\
$\nu$\,Cnc      & 10400 & 3.60 &$\leq5$ &0.84&$\leq7.94$ \\
$\kappa$\,Cnc   & 13500 & 3.80 & 38 & 0.69 & 8.08 \\
HR\,4072        & 10500 & 3.95 & 4.1& 0.87 & 7.96 \\
$\chi$\,Lup     & 10750 & 4.00 & 4.1& 0.87 & 7.87 \\
$\iota$\,CrB    & 11000 & 4.00 & 3.7& 0.85 & 7.67 \\
$\upsilon$\,Her & 12000 & 3.80 &$\leq1.0$&0.80&$\leq6.53$ \\
$\phi$\,Her 	& 11650 & 4.00 & 2.0 & 0.83 & 7.08 \\
HR\,6997        & 14500 & 3.90 & 37 & 0.67 & 7.82 \\
112\,Her        & 13100 & 4.10 & 7.1 & 0.77 & 7.28 \\
HR\,7143        & 12100 & 4.00 &$\leq3$&0.81&$\leq7.10$ \\
HR\,7361   	& 13650 & 3.55 & 40 & 0.65 & 7.94 \\
46\,Aql         & 13000 & 3.65 & 9 &  0.72 & 7.23 \\
HR\,7664	& 13200 & 3.60 & 29 & 0.68 & 7.85 \\

\hline
\end{tabular}

\label{table:nlte_abunds}

\end{table}

\subsection{Neon abundances from Non-LTE calculations}
\label{nlte_abund}

\begin{figure}
\begin{center}
\epsfig{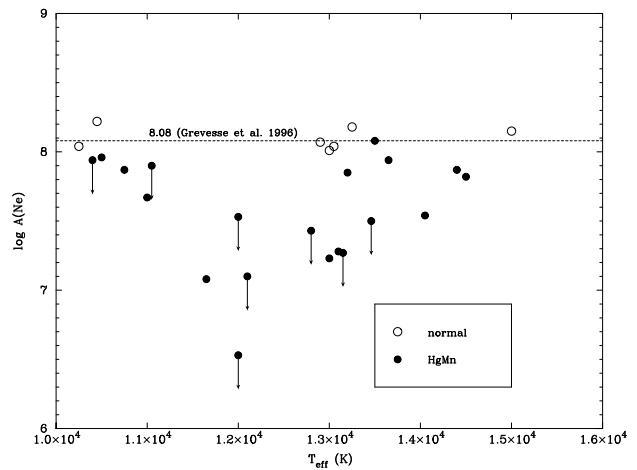}
\caption{Abundances of Ne in normal stars (open circles) and HgMn stars
(filled circles).  Upper limits for abundances are indicated by arrows.  
The standard abundance for Ne on the scale $\log A({\rm H})=12.00$ is from
\citet{grev96}.} 
\label{fig:nlte_abunds} 
\end{center}
\end{figure}

One may obtain a `corrected' LTE equivalent width from $R W_\lambda({\rm
obs})$ and analyse the corrected width by using a standard LTE approach
including the appropriate microturbulence in a fully line-blanketed case.  
We used {\sc uclsyn} to calculate the abundance of Ne from equivalent
widths given in Table~\ref{table:ewidths} after scaling by $R$ from
Table~\ref{table:ratios}.  The microturbulence parameters $\xi$ were taken
from Table~\ref{table:stars}. The results are shown in
Table~\ref{table:nlte_abunds} and plotted as a function of \teff\ in
Figure~\ref{fig:nlte_abunds}.

Our results yield for normal and superficially-normal stars a mean
abundance $\log A ({\rm Ne}) = 8.10\pm0.03$ relative to 12.00 for H.  
This is in excellent agreement with the standard abundance of Ne, 8.08,
given by \citet{grev96}, which is essentially identical to the value
(8.09) in the earlier compilation of \citet{ag89}. The standard abundance
is partly based on local galactic values (stars and nebulae), and on
application of a well-determined correction to solar wind and solar
energetic particle values, as well as spectroscopy of solar prominences.  
We take this agreement as confirmation that our ratio method for LTE/NLTE
equivalent width scaling works well in the \teff\ range of HgMn stars.

The HgMn stars are, with only one exception ($\kappa$\,Cnc), deficient in
Ne, although the deficits in a few cases are only marginal ($0.1-0.2$
dex). In many cases we are able only to establish upper limits for Ne
abundances.  The most extreme case is $\upsilon$\,Her, for which we have
particularly good spectra and were able to establish an upper limit 1.5
dex below the solar abundance.  There is no case in our sample where Ne
has an abundance greater than the standard value.  The results of Adelman
\& Pintado (2000) can also be analysed using our method.

\section{Discussion}
\label{disc}

The error found for the average abundance for the normal stars is based on
the scatter in the results.  For comparison, following procedures adopted
in previous papers in this series on HgMn stars (e.g.,
\citealt{sd93,jda99}) we propagate uncertainties in adopted estimates of
the errors on each parameter as follows: $\pm$0.25\,dex in \logg,
$\pm$250\,K in \teff, $\pm$0.5\,km\,s$^{-1}$ in the microturbulence
($\xi$), and $\pm$5 percent in $W_{\lambda}$. Propagating these errors
through the `corrected' LTE analysis used above for the \nei $\lambda$6402
line, using a model atmosphere at \teff\ = 13\,000\,K, \logg\ = 4.0,
$\xi$=1, leads to the following representative errors in the derived Ne
abundances: $\pm$0.10\,dex (\logg); $\pm$0.08\,dex (\teff); $\pm$0.01\,dex
($\xi$); $\pm$0.04\,dex ($W_{\lambda}$) at standard abundance
($W_{\lambda}=27$\,m\AA), and $\pm$0.10\,dex (\logg); $\pm$0.07\,dex
(\teff); $\pm$0.01\,dex ($\xi$);  $\pm$0.03\,dex ($W_{\lambda}$) with neon
underabundant by 0.5 dex ($W_{\lambda}=13$\,m\AA). These are very similar
ranges; the combined expected error for one measurement is $\pm0.13$\,dex.
The s.d. for the normal stars is $\pm0.08$\,dex.  This difference may
reflect overestimates in some of the above factors (especially $\Delta\log
g$) that comprise the estimated errors, but the two estimates are not in
serious disagreement.

We have implicitly conducted all our analyses under the assumption of a
homogeneous depth distribution of neon in the photospheres of the HgMn
stars. It now seems well-founded to conclude that in many HgMn stars the
neon atoms may not be distributed with a constant fraction vs. optical
depth, due to gravitational settling, but our results offer no method of
distinguishing clearly between a uniform depletion in the line forming
region and an inhomogeneous distribution in which the total number of
absorbers is about the same.  Given the scatter in abundance from star to
star, line strengths alone will be inadequate to prove the point one way
or the other.

One should consider the question of whether or not we may lump together
the normal and superficially normal stars.  The `superficially normal'
stars listed in Tables~\ref{table:stars},~\ref{table:ewidths} and
\ref{table:nlte_abunds} were originally described as such by
\citet{cow80}, owing to their relatively sharp lines, but subsequent
investigations by \citet{sd90,sd93} and \citet{kcs93,kcs94,kcs96,kcs97}
have shown that HR\,7338, HR\,7878, and 21\,Peg have abundances which are
not distinguishable from normal stars for C, N, Cr$-$Ni, Mg, Al, Si, Cu,
Zn, Ga and Hg. \citeauthor{cow80} thought that HR\,7878 and HR\,7338 were
normal stars with no trace of peculiarity, although he suspected that
21\,Peg might be related to early Am stars due to a weaker than expected
Sc~{\sc{ii}} line, and \citet{sad80} found that both 21\,Peg and HR\,7338
may have `hot Am' characteristics, such as mild Ba and Y enhancements. One
of these, 21\,Peg, is listed as Hg peculiar class by \citet{renson}
without a reference cited, but Smith's (\citeyear{kcs97}) study of
Hg~{\sc{ii}} lines showed that the Hg abundance is effectively
indistinguishable from that found in normal stars.  \cite{land98} also
describes it as a normal star.  In what follows we assume that 21\,Peg is
a normal B9.5\,V star and that HR\,7338 is also normal.  In this work we
therefore feel justified in including these three stars with the normal
stars in Table~\ref{table:nlte_abunds} and Fig.~\ref{fig:nlte_abunds}.

The equivalent width measures of \citet{adel00} for 3\,Cen\,A,
$\kappa$\,Cnc, and HR\,7245 can be used with the results of
Section~\ref{section:NLTE} to derive approximate NLTE abundances for neon.  
For the first two of these stars, those authors give only the equivalent
width of $\lambda$5852.49 (although their table headings say 5842.49,
which is evidently a misprint).  We assume, based on the discussion of
Section~\ref{sub:ew_LTEabunds}, that the correction factor $R$ may be
taken to be about the same as for $\lambda$6402, and we further assume
that we can extrapolate the correction factor to \teff\ = 17\,500 for
3\,Cen\,A, for which we assume $R=0.59$.  This well-known peculiar He-weak
star has some characteristics similar to HgMn stars.  We find a
near-standard Ne abundance of 8.17, while $\kappa$\,Cnc has a slight
overabundance (8.32), and the HgMn star HR\,7245, which has a measured
equivalent width of 8 m\AA, has a low abundance (7.32) similar to that of
112\,Her.  Given that our assumptions above could be subject to some
uncertainty, at this stage we would not wish to conclude much more than
that the abundance of neon seems consistent with the standard value in
3\,Cen\,A and $\kappa$\,Cnc, but in the case of HR\,7245 we are probably
on firm ground in assigning a very low abundance of neon.

We explored briefly the question of whether the Ne abundances in HgMn
stars depend on atmospheric parameters. It seems that largest anomalies
(underabundances) are generally observed in the middle of the temperature
range of HgMn stars ($11500<\teff<13000$\,K; see
Figure~\ref{fig:nlte_abunds}). No apparent correlation with the surface
gravity can be seen in Table~\ref{table:nlte_abunds}.  It is not possible
to draw any conclusions about dependences on rotational velocity as we
have chosen to work with a selected sample of HgMn stars with fairly small
\vsini\ in order to ensure accurate abundance determinations.

\section{Conclusions}

We have measured the equivalent widths (or upper limits) of several \nei\
lines in the spectra of 11 normal late B stars and 21 HgMn stars in the
same \teff\ range. These lines were selected after a search for lines
which were well placed in \'{e}chelle orders in the HES and which appeared
not to have any significant blending features in the sample of stars
studied.  When analysed using LTE methods in fully line-blanketed
atmospheres, there is a steady increase with \teff\ in the apparent
abundance above the standard value of $\log A({\rm Ne})=8.08$.  It is
apparent from previous studies that this is due to NLTE effects.  We note
that the strongest line of Ne in the red region, $\lambda$6402, and the 6
other lines studied, generally give concordant LTE abundances, suggesting
that they are affected by NLTE effects by roughly the same amount.  These
lines may be of use in future investigations, provided further
observations and NLTE calculations are made.

We undertook a detailed NLTE analysis of $\lambda$6402 by use of a full
analysis including NLTE for H {\sc i} and \nei.  We confirm earlier
studies by obtaining very similar results when similar inputs are used,
and find that the ratio of the NLTE and LTE equivalent widths calculated
for a given NLTE model atmosphere is a slowly varying function of \teff\
and surface gravity.  This ratio only slightly depends on the actual
abundance of Ne, and is also very insensitive to the microturbulence, so
it becomes possible to interpolate, in a table of the ratio $R$, the
`correction factor' by which an observed equivalent width must be scaled
in order to produce the NLTE abundance from a much easier LTE analysis.

The normal stars in our sample yield a mean logarithmic Ne abundance of
8.10, well within our formal mean error of $\pm0.03$ of the standard
value, 8.08, given by \citet{grev96}.  This gives us additional confidence
that our models, and the ratio method of using LTE calculations as an
interpolation device, work satisfactorally for late B stars.  The
smallness of the scatter (s.d. $\pm0.08$) for the individual stars
suggests that the error budget in Section~\ref{disc} is rather
conservative.

It is clear from our results for the HgMn stars that the abundances of Ne
range from standard abundance or slightly below, to extreme deficiencies
of an order of magnitude or more.  In several cases we have obtained only
upper limits, and additional observations of very high quality would be
needed in order to attempt to detect the weak Ne lines in these stars.
There is a tendency for the Ne abundance to be smallest in the middle of
the HgMn effective temperature range, but there is no dependence on
surface gravity.  There is not a single confirmed case in which Ne has an
enhanced abundance, which is strong evidence for the absence in HgMn stars
of nonturbulent stellar winds (i.e. hydrogen mass loss rate must be $<
10^{-14}M_{\sun}$\,yr$^{-1}$) that might compete with radiative atomic
diffusion and produce accumulations of light elements in the photosphere,
as suggested by \citet{landolvau}. That such winds might exist was studied
by \citet{babmich}, \citet{babel92} and \citet{krtkub}.

\section{Acknowledgements} 

J. Budaj gratefully acknowledges the support of The Royal Society for a
Royal Society/NATO Fellowship (ref 98B) and further support from VEGA
Grant No. 7107 from the Slovak Academy of Sciences. MMD is grateful to the
Director of the University of California Observatories, Prof.\ J.S.
Miller, for allocating time for his Guest Observer programme.  We
gratefully acknowledge the work of C. M. Jomaron and Mr. Christopher Jones
for their assistance in this work.  Thanks also to the technical and
service staff at Lick Observatory, and especially to Tony Misch, for their
efforts on our behalf. NSO/Kitt Peak FTS data used for the measurement of
parasitic light were produced by NSF/NOAO. We are grateful to I. Hubeny
and J. Krti\v{c}ka for numerous discussions. Research on chemically
peculiar stars at UCL is supported by PPARC grant GR/K58500. C. Jones was
supported by an Undergraduate Research Bursary from the Nuffield
Foundation (NUB-URB98).  This work was also supported by the PPARC PATT
Rolling Grant GR/K60107 to UCL, for travel to telescopes.


\begin{thebibliography}{}

\bibitem[\protect\citeauthoryear{Adelman}{1988a}]{adel88a}
Adelman S.J., 1988a, MNRAS, 235, 749

\bibitem[\protect\citeauthoryear{Adelman}{1988b}]{adel88b}  
Adelman S.J., 1988b, MNRAS, 235, 763

\bibitem[\protect\citeauthoryear{Adelman}{1989}]{ad89}
Adelman S.J., 1989, MNRAS, 239, 487

\bibitem[\protect\citeauthoryear{Adelman \& Fuhr}{1985}]{adfuhr85}
Adelman S.J., Fuhr J.R., 1985, A\&A, 152, 434

\bibitem[\protect\citeauthoryear{Adelman \& Pintado}{2000}]{adel00}
Adelman S.J., Pintado O.I., 2000, A\&A, 354, 899

\bibitem[\protect\citeauthoryear{Adelman, Philip \& Adelman}{Adelman et
al.}{1996}]{ad96}
Adelman S.J., Philip A.G.D., Adelman C.J., 1996, MNRAS, 282, 953

\bibitem[\protect\citeauthoryear{Allen}{1998}]{csa98}
Allen C.S., 1998, PhD thesis, University of London
\newline URL=http://www.ulo.ucl.ac.uk/ulo\_comms\\/80/index.html

\bibitem[\protect\citeauthoryear{Anders \& Grevesse}{1989}]{ag89}
Anders E., Grevesse N., 1989, Geochim. Cosmochim. Acta, 53, 197

\bibitem[\protect\citeauthoryear{Auer \& Mihalas}{1973}]{auer73}
Auer L.H., Mihalas D., 1973, ApJ, 184, 15

\bibitem[\protect\citeauthoryear{Babel}{1992}]{babel92} Babel J., 1992,
A\&A, 258, 449

\bibitem[\protect\citeauthoryear{Babel \& Michaud}{1991}]{babmich} Babel
J., Michaud G., 1991, ApJ, 366, 560

\bibitem[\protect\citeauthoryear{Batten, Fletcher \& MacCarthy}{Batten et
al.}{1989}]{batten89}
Batten A.H., Fletcher J.M., MacCarthy D.G., 1989, Publs. DAO, 17, 1

\bibitem[\protect\citeauthoryear{Bohlender, Dworetsky \&
Jomaron}{Bohlender et al.}{1998}]{bohl}
Bohlender D.A., Dworetsky M.M., Jomaron C.M., 1998, ApJ, 504, 533

\bibitem[\protect\citeauthoryear{Budaj \& Dworetsky}{2000}]{budw00}
Budaj J., Dworetsky M. M., 2000, in preparation

\bibitem[\protect\citeauthoryear{Churchill}{1995}]{churchill}
Churchill C.W., 1995, Lick Obs. Tech. Rep. No. 74

\bibitem[\protect\citeauthoryear{Cowley}{1980}]{cow80}
Cowley C.R., 1980, PASP, 92, 159

\bibitem[\protect\citeauthoryear{Cunto et al.}{1993}]{cunto}
Cunto W.C., Mendoza C., Ochsenbein F., Zeippen C.J., 1993, A\&A, 275, L5

\bibitem[\protect\citeauthoryear{del Val, Aparicio \& Mar}{del Val et
al.}{1999}]{delval}
del Val J.A., Aparicio J.A., Mar A., 1999, ApJ, 513, 535

\bibitem[\protect\citeauthoryear{Dworetsky \& Budaj}{2000}]{dwbu2000}
Dworetsky M. M., Budaj J., 2000, Comm. Univ. London Observatory. No. 81
\newline URL=http://www.ulo.ucl.ac.uk/ulo\_comms\\/81/index.html

\bibitem[\protect\citeauthoryear{Dworetsky, Jomaron \& Smith}{Dworetsky et
al.}{1998}]{djs98} Dworetsky M.M., Jomaron C.M., Smith C.A., 1998, A\&A,
333, 665

\bibitem[\protect\citeauthoryear{ESA}{1997}]{ESA1997}
ESA (European Space Agency), 1997, The Hipparcos and Tycho Catalogues, ESA
SP-1200

\bibitem[\protect\citeauthoryear{Gigas}{1986}]{gigas}
Gigas D., 1986, A\&A, 165, 170

\bibitem[\protect\citeauthoryear{Grevesse, Noels \& Sauval}{Grevesse et
al.}{1996}]{grev96}
Grevesse N., Noels A., Sauval A. J., 1996, in Holt S. \& Sonneborn G.,
eds, Cosmic Abundances, PASPC No. 99, p. 117 

\bibitem[\protect\citeauthoryear{Griem}{1974}]{griem74} Griem H. R., 1974,
Spectral Line Broadening by Plasmas, Academic Press, New York, Appendix
IVa

\bibitem[\protect\citeauthoryear{Harman}{1997}]{harman}
Harman D.J., 1997, MSci Project Report, University College London

\bibitem[\protect\citeauthoryear{Hartmetz \& Schmoranzer}{1984}]{hartmetz}
Hartmetz P., Schmoranzer H., 1984, Z. Phys. A, 317, 1

\bibitem[\protect\citeauthoryear{Hibbert \& Scott}{1994}]{hibb}
Hibbert A., Scott M.P., 1994, J. Phys. B 27, 1315

\bibitem[\protect\citeauthoryear{Hoffleit \& Warren}{1991}]{bsc91}
Hoffleit D., Warren Jr. W.H., 1991, Bright Star Catalogue, 5th Rev. Ed., 
ftp://cdsarc.u-strasbg.fr/cats/V/50 

\bibitem[\protect\citeauthoryear{Howarth et al.}{1998}]{dipso} Howarth
I.D., Murray J., Mills D., Berry D.S., 1998, Starlink User Note 50,
Rutherford Appleton Lab./CCLRC

\bibitem[\protect\citeauthoryear{Hubeny}{1981}]{hub81}
Hubeny I. 1981, A\&A, 98, 96

\bibitem[\protect\citeauthoryear{Hubeny}{1988}]{hub88}
Hubeny I., 1988, Comput. Phys. Comm., 52, 103

\bibitem[\protect\citeauthoryear{Hubeny \& Lanz}{1992}]{hula92}
Hubeny I., Lanz T., 1992, A\&A, 262, 501

\bibitem[\protect\citeauthoryear{Hubeny \& Lanz}{1995}]{hula95}
Hubeny I., Lanz T. 1995, ApJ, 439, 875

\bibitem[\protect\citeauthoryear{Hubeny, Lanz \& Jeffery}{Hubeny et
al.}{1995}]{hulajef95}
Hubeny I., Lanz T., Jeffery C.S., 1995, Tlusty \& Synspec - A User's Guide

\bibitem[\protect\citeauthoryear{Hubrig \& Launhardt}{1993}]{hl93} 
Hubrig S., Launhardt R., 1993, in Dworetsky M.M., Castelli F., Faraggianna
R., eds, Peculiar versus Normal Phenomena in A-Type and Related Stars.
PASPC No.  44, p. 350

\bibitem[\protect\citeauthoryear{Jomaron, Dworetsky \& Allen}{Jomaron et
al.}{1999}]{jda99} Jomaron C.M., Dworetsky M.M., Allen C.S., 1999, MNRAS,
303, 555

\bibitem[\protect\citeauthoryear{Krti\v{c}ka \& Kub\'{a}t}{2000}]{krtkub}
Krti\v{c}ka J., Kub\'{a}t J., 2000, A\&A, in press

\bibitem[\protect\citeauthoryear{Kurucz}{1990}]{cd23}
Kurucz R.L., 1990, Trans. IAU, XXB, 168 (CD-ROM 23)

\bibitem[\protect\citeauthoryear{Kurucz}{1993}]{cd13}
Kurucz R.L., 1993, ATLAS9 Stellar Atmosphere Programs and 2 km s$^{-1}$
Grid (CD-ROM 13)

\bibitem[\protect\citeauthoryear{Kurucz et al.}{1984}]{kpno}
Kurucz R.L., Furenlid I., Brault J., Testerman L., 1984, Solar Flux
Atlas from 296 nm to 1300 nm (NSO Atlas No 1)

\bibitem[\protect\citeauthoryear{Landstreet}{1998}]{land98}
Landstreet J. D., 1998, A\&A, 338, 1041

\bibitem[\protect\citeauthoryear{Landstreet, Dolez \& Vauclair}{Landstreet
et al.}{1998}]{landolvau} 
Landstreet J. D., Dolez N., Vauclair S., 1998, A\&A, 333, 977

\bibitem[\protect\citeauthoryear{Lanz}{1995}]{l95}
Lanz T., 1995, In: Adelman S.J., Wiese W.L., eds,
Astrophysical Applications of Powerful New Databases, PASPC No. 78, p. 423

\bibitem[\protect\citeauthoryear{Michaud}{1970}]{m70}
Michaud G., 1970, ApJ, 160, 641

\bibitem[\protect\citeauthoryear{Michaud}{1986}]{m86}
Michaud G., 1986, In Cowley C.R., Dworetsky M.M., M\'{e}gessier C.,
eds, Proc. IAU Coll. 90, Upper Main Sequence Stars with Anomalous
Abundances, Reidel, Dordrecht, p. 459

\bibitem[\protect\citeauthoryear{Mihalas}{1978}]{mihalas78}
Mihalas D., 1978, Stellar Atmospheres, 2nd Ed., W.H. Freeman \& Co.,
San Francisco

\bibitem[\protect\citeauthoryear{Misch}{1997}]{mischb} 
Misch A., 1997, User's Guide to the Hamil\-ton Echelle
Spec\-tro\-me\-ter\\ 
URL=http://www.ucolick.org/\verb+~+tony/instruments\\/hamspec/hamspec\_index.html

\bibitem[\protect\citeauthoryear{Moore}{1949}]{moore}
Moore C.E., 1949, Atomic Energy Levels, Vol. I, NBS Circular No. 467,
Washington, D.C.

\bibitem[\protect\citeauthoryear{Petrie}{1950}]{petrie} Petrie R.M., 1950,
Publs. DAO, 8, 319

\bibitem[\protect\citeauthoryear{Renson, Gerbaldi \& Catalano}{Renson et
al.}{1991}]{renson}
Renson P., Gerbaldi M., Catalano F. A., 1991, AAS, 89, 429

\bibitem[\protect\citeauthoryear{Ryabchikova, Zakhorova \&
Adelman}{Ryabchikova et al.}{1996}]{r96} Ryabchikova T.A., Zakhorova L.A.,
Adelman S.A., 1996, MNRAS, 283, 1115

\bibitem[\protect\citeauthoryear{Ryabchikova et al.}{1998}]{r98}
Ryabchikova T., Kotchoukhov O., Galazutdinov F., Musaev
F., Adelman S.J., 1998, Contrib. Astron. Obs. Skalnat\'{e} Pleso, 27, 258

\bibitem[\protect\citeauthoryear{Sadakane}{1980}]{sad80}
Sadakane K., 1980, PASP, 93, 587

\bibitem[\protect\citeauthoryear{Seaton}{1998}]{mjs}
Seaton M.J., 1998, J Phys B, 31, 5315

\bibitem[\protect\citeauthoryear{Sigut}{1999}]{sigut}
Sigut T. A. A., 1999, ApJ, 519, 303

\bibitem[\protect\citeauthoryear{Smith}{1992}]{kcs92}
Smith K.C., 1992, PhD thesis, University of London

\bibitem[\protect\citeauthoryear{Smith}{1993}]{kcs93}
Smith K.C., 1993, A\&A, 276, 393

\bibitem[\protect\citeauthoryear{Smith}{1994}]{kcs94}
Smith K.C., 1994, A\&A, 291, 521

\bibitem[\protect\citeauthoryear{Smith}{1996}]{kcs96}
Smith K.C., 1996, A\&A, 305, 902

\bibitem[\protect\citeauthoryear{Smith}{1997}]{kcs97}
Smith K.C., 1997, A\&A, 319, 928

\bibitem[\protect\citeauthoryear{Smith \& Dworetsky}{1988}]{sd88}
Smith K.C., Dworetsky M.M., 1988, In: Adelman S.J., Lanz T., eds,
Elemental Abundance Analyses.  Institut d'Astronomie de l'Univ.  de
Lausanne, Switzerland, p. 32

\bibitem[\protect\citeauthoryear{Smith \& Dworetsky}{1990}]{sd90}
Smith K. C., Dworetsky M. M., 1990, in Rolfe, E. J., Evolution in
Astrophysics, ESA SP-310, p. 279

\bibitem[\protect\citeauthoryear{Smith \& Dworetsky}{1993}]{sd93}
Smith K.C., Dworetsky M.M., 1993, A\&A, 274, 335

\bibitem[\protect\citeauthoryear{Valdes}{1990}]{valdes}
Valdes F., 1990, The IRAF APEXTRACT Package, 
\newline URL=ftp://iraf.tuc.noao.edu/iraf/docs/apex.ps

\bibitem[\protect\citeauthoryear{Varosi et al.}{1995}]{varosi}
Varosi F., Lanz T., Dekoter A., Hubeny I., Heap S., 1995, MODION (NASA
Goddard SFC)
\newline URL=ftp://idlastro.gsfc.nasa.gov/pub/contrib\\/varosi/modion/

\bibitem[\protect\citeauthoryear{Vauclair \& Vauclair}{1982}]{vau82}
Vauclair S., Vauclair G., 1982, ARA\&A, 20, 37

\bibitem[\protect\citeauthoryear{Vogt}{1987}]{vogt}
Vogt S., 1987, PASP, 99, 1214

\bibitem[\protect\citeauthoryear{Wahlgren, Adelman \& Robinson}{Wahlgren
et al.}{1994}]{wahl94} 
Wahlgren G.M., Adelman S.J., Robinson R.D., 1994, Ap.J., 434, 349

\bibitem[\protect\citeauthoryear{Warner}{1967}]{W67}
Warner B., 1967, MNRAS, 136, 381

\bibitem[\protect\citeauthoryear{Wiese, Smith \& Glennon}{Wiese et
al.}{1966}]{wsg66}
Wiese W.L., Smith M.W., Glennon B.M., 1966, Atomic Transition
Probabilities, Vol. I, NSRDS-NBS 4

\end{thebibliography}
\end{document}